\documentclass[12pt]{iopart}
\usepackage{graphicx}

\begin{document}
\title[Reconstruction of a single-active-electron potential...]{Reconstruction of a single-active-electron potential from electron momentum distribution produced by strong-field ionization using optimization technique}

\author{N I Shvetsov-Shilovski}

\address{Institut f\"{u}r Theoretische Physik, Leibniz Universit\"{a}t Hannover, D-30167, Hannover, Germany}

\ead{n79@narod.ru}

\begin{abstract}
We present a method for retrieving of single-active electron potential in an atom or molecule from a given momentum distribution of photoelectrons ionized by a strong laser field. In this method the potential varying within certain limits is found as the result of the optimization procedure aimed at reproducing the given momentum distribution. The optimization using numerical solution of the time-dependent Schrodinger equation for ionization of a model one-dimensional atom shows the good accuracy of the potential reconstruction method. 
This applies to different ways used for representing of the potential under reconstruction, including a parametrization and determination of the potential by specifying its values on a spatial grid. 
\end{abstract}


\submitto{\jpb}
\maketitle

\section{Introduction}

Ionization of molecules by strong laser pulses has been attracting considerable attention both in experiment and theory, see \cite{Marangos2004, Bandrauk2003} for recent reviews. The reason for this is that the accurate description of ionization step is necessary for understanding of a variety of phenomena arising from the interaction of strong laser pulses with molecules. These phenomena include above-threshold ionization (ATI), formation of the high-energy plateau in the ATI spectrum (high-order ATI), generation of high-order harmonics, nonsequential double ionization, etc (see \cite{BeckerRev2002, MilosevicRev2003, FaisalRev2005, FariaRev2011} for reviews). Among the main theoretical approaches used to describe ionization of atoms by strong laser pulses are the direct numerical solution of the time-dependent Schr\"{o}dinger equation (TDSE) (see, e.g., \cite{Muller1999, Bauer2006, Madsen2007, Tong2017}), the strong-field approximation \cite{Keldysh1964, Faisal1973, Reiss1980}, and the semiclassical models using classical mechanics to describe the motion of an electron after it has been promoted to the continuum. The two-step \cite{Linden1988, Gallagher1988, Corkum1989} and the three-step \cite{Kulander_Schafer1993, Corkum1993} models are the widely known examples of semiclassical approaches. 

All these theoretical methods have been generalized to the molecular case. Obviously, it is more difficult to describe the strong-field ionization of a molecule than this process in an atom. Indeed, due to the nuclear motion a molecule has extra degrees of freedom as compared to the atomic case. These additional degrees of freedom lead to a necessity of taking into account the corresponding time scales. Furthermore, electronic orbitals in molecules often have complicated shapes. In principle, the direct numerical solution of the TDSE allows to fully take into account all the features of the molecular ionization. However, this numerical solution of the three-dimensional (3D) TDSE for a molecule interacting with a strong laser pulse is a very complicated task. It is possible to solve the 3D TDSE only for the simplest molecules selecting the most relevant degrees of freedom \cite{Palacios2006, Saenz2014}. Therefore, a number of approximations has to be used to solve the TDSE for strong-field ionization of a molecule. Usually it is assumed that the positions of the atomic nuclei are fixed. Simultaneously, the single-active electron (SAE) approximation \cite{Kulander1988, Kulander1991} is applied. Within the SAE the strong-field ionization of an atom or molecule is described as an interaction of only one active electron with the laser radiation. This active electron moves in the combined effective potential and the electric field of the laser pulse. Therefore, it is important to obtain suitable SAE potentials for various molecules. However, the calculation of the SAE potential is a non-trivial task.  

The potential that describes the interaction of one active electron with the frozen core can be approximated as $V\left(\mathbf{r}\right)=V_{en}\left(\mathbf{r}\right)+V_{ee}^{cl}\left(\mathbf{r}\right)+V_{xc}\left(\mathbf{r}\right)$, where $V_{en}\left(\mathbf{r}\right)$ is the electron-electron nuclear interaction, $V_{ee}^{cl}\left(\mathbf{r}\right)$ is the Hartree electron-electron repulsion, and $V_{xc}\left(\mathbf{r}\right)$ is the exchange-correlation potential. The occupied orbitals and therefore the Hartree term $V_{ee}^{cl}\left(\mathbf{r}\right)$ can be obtained using the standard quantum chemistry packages: GAMESS \cite{Gamess}, Gaussian \cite{Gaussian}, etc. In exchange-only calculations, the exchange-correlation potential can be approximated by the local density approximation \cite{KohnSham1965}. To ensure that the resulting SAE potential has Coulomb behavior at large distances, a gradient correction to the local density \cite{vanLeeuwen1994} can be applied. The approach described here was used in \cite{Madsen2010, Lin2010} to calculate the potentials for some selected simple molecules.

However, simpler model potentials are used in many studies applying the solution of the TDSE for molecules. Some of these model potentials are constructed as simple combinations of a few Coulomb soft-core atomic potentials with the centers at the atomic nuclei, whereas the others rely on more complicated expressions (see, e.g., \cite{Miller1974, Kastner2010, Atabek2012, Ding2016}). Usually the model potential depends on one or few parameters. While one parameter of the model potential allows to reproduce only one bound-state energy, the presence of several parameters makes it possible to recover the predefined energies of a few different bound states. Usually the parameters of a model potential are adjusted using an optimization technique. This suggests to explore the possibility of obtaining the SAE potential as a result of some optimization procedure. It is well known that the potential experienced by the ionized electron is encoded in the measured photoelectron momentum distribution (PMD). Therefore, it should be possible to optimize the unknown SAE potential in order to reproduce a given PMD generated by strong-field ionization of an atom or molecule. 

In this paper we develop such an approach to retrieval of the SAE potentials. We assume that the given momentum distribution is measured in an experiment and refer to it as the target PMD (TPMD). The TPMD is considered as a goal that is to be achieved by an optimization algorithm considering the unknown values of the potential (or expansion coefficients of the potential in a given basis) as parameters that are optimized. It should be stressed that there is a substantial difference between the approach proposed here and the quantum optimal control theory (QOCT), see \cite{WerschnikGross2007, Rabitz2010} for reviews. The latter provides a powerful theoretical approach to the optimization and control of various quantum phenomena, including those in strong laser fields (see, e.g., \cite{Castro2009, Solanpaa2014, Shvetsov2015, Koch2016}). Indeed, the QOCT treats the electric field of the laser pulse as a control function, whereas the effective potential experienced by an electron is to be predefined and does not change in optimization.

Any optimization requires a specification of a measure that allows to identify whether the optimization target is achieved. Since a PMD is a picture, it is natural to specify such a measure using tools employed in image analysis and pattern recognition. Therefore, the measures applied to compare different digital images or videos (see, e.g. \cite{Wang2004, Buldas2013}) can be used to estimate the similarity of the two PMD's: the result of current iteration of an optimization algorithm and the target distribution. However, a valid choice of the specific measure and its application to comparison of the PMD's require thorough studies. For this reason, we leave the application of image recognition tools for future investigations.
In this paper we retrieve the SAE potential from momentum distribution produced in ionization of a one-dimensional (1D) model atom. The 1D momentum distribution is a function of only one variable (momentum component along one spatial axis). Therefore, the widely-known measures used in variation calculus and functional analysis (see, e.g., \cite{KolmogorovFomin}) can be applied for comparison of different PMD's.

It was shown in the studies of the ATI process that the vast majority of the photoelectrons do not recollide with their parent ions.
These electrons are referred to as direct electrons. They are detected with the energies below $2U_{p}$, where $U_{p}=F_{0}^{2}/\omega^2$ is the ponderomotive energy. Here $F_{0}$ is the field strength and $\omega$ is the angular frequency (atomic units are used throughout the paper). There are also rescattered electrons that due to oscillations of the laser field come back to their parent ions and rescatter on them by large angles. The rescattered electrons form the high-energy plateau in the ATI spectrum, i.e., they are responsible for the high-order ATI. In our optimization-based approach only momentum distributions of the direct electrons are used. The optimization technique relying on the distributions of rescattered electrons will be the subject of further studies.  

The paper is organized as follows. In Sec.~II we briefly discuss our approach to solve the 1D TDSE, measures used to compare electron momentum distributions, and derivative-free optimization algorithms. In Sec.~III we apply our method to retrieve the soft-core Coulomb potential from the PMD's produced by ionization of a 1D model atom. We test our approach for two different ways of representing the unknown potential and discuss various optimization strategies. The conclusions and outlook are given in Sec. IV.

\section{Comparison of electron momentum distributions and optimization}
\subsection{Numerical solution of the time-dependent Schr\"{o}dinger equation}
We define a few-cycle laser pulse linearly polarized along the $x$-axis by specifying its vector-potential:
\begin{equation}
\mathbf{A}\left(t\right)=\left(-1\right)^{n_p+1}\frac{F_0}{\omega}\sin^2\left(\frac{\omega t}{2n_p}\right)\sin\left(\omega t +\varphi\right)\mathbf{e}_{x}, 
\label{vecpot}
\end{equation}
where $n_p$ is number of optical cycles within the pulse, $\varphi$ is the carrier envelope phase, $\mathbf{e}_x$ is a unit vector, and the laser pulse is present between $t=0$ and $T=\left(2\pi/\omega\right)n_{p}$. The electric field can be calculated from vector-potential (\ref{vecpot}) by $\mathbf{F}=-d\mathbf{A}/dt$. 
In the velocity gauge, the 1D TDSE for an electron interacting with the lase pulse is given by
\begin{equation}
i\frac{\partial}{\partial t}\Psi\left(x,t\right)=\left[\frac{1}{2}\left(-i\frac{\partial}{\partial x}+A_{x}\left(t\right)\right)^2+V\left(x\right)\right]\Psi\left(x,t\right),
\label{tdse}
\end{equation}
where $\Psi\left(x,t\right)$ is the time-dependent wave function in coordinate representation and $V\left(x\right)$ is the SAE potential. In the absence of the electric field, the time-independent Schrodinger equation reads as
\begin{equation}
\left[-\frac{1}{2}\frac{d^2}{dx^2}+V\left(x\right)\right]\Psi\left(x\right)=E\Psi\left(x\right),
\label{tise}
\end{equation}
where $\Psi\left(x\right)$ and $E$ are the eigenfunction and the corresponding energy eigenvalue, respectively. The eigenvalue problem (\ref{tise}) is solved on a grid using the three-step formula to approximate the second derivative. Hence, the diagonalization routine developed for sparse matrices \cite{Anderson1999} can be used. Alternatively, the ground state and the first few excited states can be found by imaginary time propagation (see, e.g., \cite{Kjeldsen2007, BauerBook}). The computational box is centered at $x=0$ and extends to $\pm x_{max}$, i.e., $x\in\left[-x_{max},x_{max}\right]$. Typically, we set $x_{max}=250.0$~a.u and use a grid consisting of $N=4096$ points, what corresponds to the grid spacing $dx=0.1225$~a.u. 

The well-known split-operator method \cite{Feit1982} is used to solve the TDSE (\ref{tdse}). The time step is $\Delta t=0.055$~a.u. We prevent unphysical reflections from the boundary of the grid by using absorbing boundaries, i.e., at every step of the time propagation we multiply the wave function in the region $\left|x\right|>x_{b}$ by the mask:
\begin{equation}
M\left(x\right)=\cos^{1/6}\left[\frac{\pi\left(\left|x\right|-x_{b}\right)}{2\left(x_{max}-x_{b}\right)}\right]
\label{mask}
\end{equation}
with $x_b=3x_{max}/4$. Hence, $x=\pm x_{b}$ correspond to the internal boundaries of the absorbing regions. As the result, at every time step the part of the wave function in the mask region is absorbed without any effect on the $\left|x\right|<x_{b}$ domain. We calculate the electron momentum distributions using the mask method \cite{Tong2006}. 

\subsection{Comparison of electron momentum distributions and optimization algorithms}
The optimization procedure developed here is based on comparison of the 1D electron momentum distributions, which are functions of one variable. The following metrics are widely used to calculate the distance between continuous functions $f\left(x\right)$ and $g\left(x\right)$ defined for $x\in\left[a, b\right]$:
\begin{equation}
\rho_{1}\left[f\left(x\right),g\left(x\right)\right]=\max_{x\in\left[a,b\right]}\left|f\left(x\right)-g\left(x\right)\right|
\label{metr1}
\end{equation}
and
\begin{equation}
\rho_{2}\left[f\left(x\right),g\left(x\right)\right]=\left\{\int_{a}^{b}dx\left[f\left(x\right)-g\left(x\right)\right]^{2}\right\}^{1/2},
\label{metr2}
\end{equation}
see, e.g., \cite{KolmogorovFomin} for a textbook treatment. We use metric (\ref{metr2}) as a measure of difference between two PMD's. We note that before calculating the distance (\ref{metr2}) we normalize electron momentum distributions to the total ionization yield (i.e., the area under the graph of the PMD). Therefore, our optimization procedure relies only on the shape of momentum distributions, but not on the ionization probabilities. We believe that similar approach when applied to the 3D case will help to facilitate the retrieval of the unknown potential from experimental electron momentum distributions. 

Since the derivatives of the similarity measure with respect to unknown potential values (or any other parameters used to represent the potential) can be calculated only numerically, it is not practical to use any gradient-based optimization method. Instead, it is appropriate to use a derivative-free optimization technique, see \cite{Rios2013, Wild2019} for recent reviews. We apply particle swarm optimization method \cite{Kennedy1995, Shi1998}, surrogate optimization technique \cite{Gutmann2001}, and pattern search method \cite{HookeJeeves1961, Kolda2003, Audet2003}. The MATLAB system \cite{Matlab} is used for simulations. 

\section{Results and discussion}
\subsection{Reconstruction of SAE potential on a grid}
In this work we reconstruct the soft-core Coulomb potential:
\begin{equation} 
V\left(x\right)=\frac{Z}{\sqrt{x^2+a}}
\label{potential}
\end{equation}
with $Z=1.0$ and $a=1.0$, see \cite{Eberly1988}. At first, we do not use any parametrization to represent the potential (\ref{potential}). Instead, we determine the potential that is to be retrieved by specifying its values in certain points of the $x$-axis. The potential values in any other points are found by interpolation. Here we use cubic spline interpolation \cite{Press1992}. As many other single-active electron potentials the potential $V\left(x\right)$ changes more rapidly for small values of $x$. Therefore, it is natural to use a non-uniform grid to represent the potential (\ref{potential}). 

\begin{figure}[h]
\centering
\includegraphics[width=.80\textwidth]{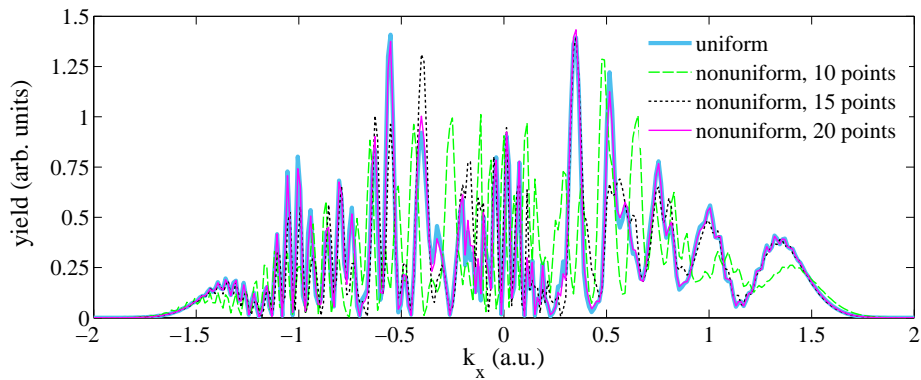}
\caption{The electron momentum distributions for ionization of a 1D atom by a laser pulse with a duration of $n_p=4$ cycles, wavelength of $800$~nm, phase $\varphi=0$, and intensity of $2.0\times10^{14}$~W/cm$^2$. The distributions are obtained from the solution of the TDSE (\ref{tdse}) with the potential (\ref{potential}) calculated on a uniform grid consisting of $N=4096$ points (thick light-blue curve), as well as with the same potential determined by its values on the non-uniform grid (\ref{nonu}) with $\gamma=20.5$ consisting of $10$ (dashed green curve), $15$ (dotted black curve), and $20$ (thin magenta curve) points.}
\label{fig1}
\end{figure}

Here we apply the following grid used for development of generalized pseudospectral methods:
\begin{equation}
x=\gamma\frac{1+x_{0}}{1-x_{0}+x_{0}^{m}},
\label{nonu}
\end{equation}
where $x_{0}^{m}=2\gamma/x_{max}$ and $\gamma$ is the mapping parameter \cite{Wang1994, Yao1993, Chu1997}. Equation (\ref{nonu}) transforms a uniform grid within the domain $x_{0}\in\left[-1.0, 1.0\right]$ to a nonuniform grid in the domain $x\in\left[0, x_{max}\right]$. The points of this grid are to be reflected with respect to $x=0$ and thus a nonuniform grid in the whole range $x\in\left[-x_{max}, x_{max}\right]$ is obtained. Then the question arises: How many points of the grid (\ref{nonu}) are required to represent the potential with sufficient accuracy? To answer this question, we choose different numbers of points for the uniform grid in the range $\left[-1.0, 1.0\right]$, find the corresponding point of the nonuniform grid (\ref{nonu}), and calculate the potential values at these points. For each number of points of the nonuniform grid we interpolate the potential $V\left(x\right)$ at every point of the dense uniform grid with $N$ points. For this interpolated potential we find energy eigenvalues and the corresponding eigenfunctions, solve the TDSE for a given laser pulse, and calculate electron momentum distributions. We next compare these PMD's with the reference momentum distribution obtained in the case that the potential (\ref{potential}) is directly calculated on the dense uniform grid consisting of $N$ points, see Figure~\ref{fig1}. It is seen that about $20$ points of the nonuniform grid (\ref{nonu}) is sufficient to reproduce the reference PMD accurately enough. Indeed, the difference between these distributions calculated in accord with the measure (\ref{metr2}) is about $0.01$. 
Taking into account that the potential $V\left(x\right)$ is an even function, the number of the points can be reduced by a factor of two, i.e., only $10$ points are needed. 

Using the nonuniform grid we first attempt to reconstruct the unknown potential using the ground state and the first excited state energies only. Such an attempt may raise questions. Indeed, it is well-known that even in the 1D case the potential cannot be unambiguously determined from one or a few energy eigenvalues. Only a symmetric {\it reflectionless} potential can be restored based on its {\it complete set} of the bound state energies \cite{Rosner1978}. However, it is evident that the optimization of a ``black-box" function that depends on $10$-$20$ parameters is a difficult numerical problem. An optimization algorithm used to solve this problem requires an initial approximation to the maximum (minimum). The success of the optimization and the convergence speed critically depend on the quality of the initial approximation. It turns out that satisfactory initial approximations can often be obtained as a result of optimization of only a few bound state energies.

\begin{figure}[h]
\centering
\includegraphics[width=0.80\textwidth,trim={5mm 0mm 20mm 0mm}]{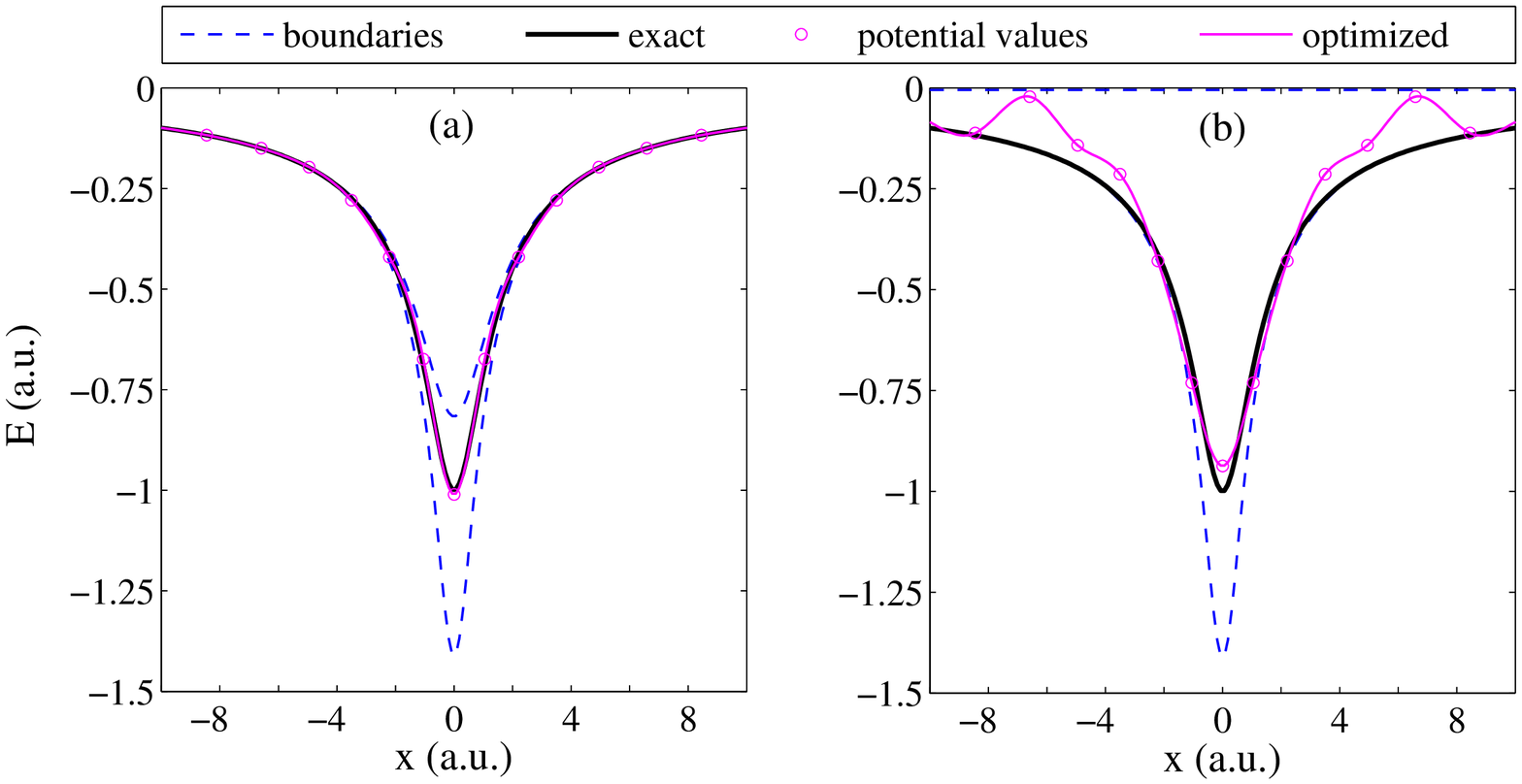} 
\caption{The values of the SAE potential on the non-uniform grid (\ref{nonu}) (magenta circles) reconstructed by optimization of the bound state and first excited states energies only, boundaries for the potential values (dashed blue curves), the potential obtained by spline interpolation based on the reconstructed values (thin magenta curve), and the soft-core Coulomb potential (thick black curve). Panel (a) shows the optimization result for the case where the optimized potential values are bounded by the potentials (\ref{sandwich}) with $Z_{1,2}=1.0$ and $a_1$ and $a_2$ equal to $0.4$ and $1.5$, respectively. Panel (b) displays the potential obtained for the optimization parameters restricted by $V_{1}\left(x\right)$ calculated from equation (\ref{sandwich}) with $Z_{1}=1.0$ and $a_1=0.4$ and $V_{2}\left(x\right)=0$. The parameters are the same as in Figure~\ref{fig1}.}
\label{fig2}
\end{figure}

The optimization methods also require the specification of the boundaries, within which the optimization parameters (in this case, the values of the potential in the grid points) can vary. We specify these ranges by sandwiching the unknown potential between the two known potential functions. As these functions we use
\begin{equation}
V_{1,2}\left(x\right)=\frac{Z_{1,2}}{\sqrt{x^2+a_{1,2}}}.
\label{sandwich}
\end{equation}
At first, we chose $Z_{1,2}=1.0$ and $a_1$ and $a_2$ equal to $0.4$ and $1.5$, respectively. Therefore, $V_1\left(x\right)$ is the lower boundary for the potential $V\left(x\right)$ that is to be retrieved, whereas $V_2\left(x\right)$ is the upper boundary. The SAE potential reconstructed by optimization the ground state energy is shown in Figure~2~(a) together with the boundaries $V_1\left(x\right)$ and $V_2\left(x\right)$. For the optimization we use the particle swarm method. It is seen that the obtained potential is in a good quantitative agreement with the one that we wanted to reconstruct. However, the situation changes dramatically, if the boundaries for the potential values under reconstruction are not as tight as in the example shown in Figure~\ref{fig2}~(a).
In Figure~\ref{fig2}~(b) we display \textit{one} of the potentials that can be obtained by optimization of the bound state energy in the case where the lower boundary for the potential values is again given by $V_1\left(x\right)$, but the upper boundary is chosen to be zero: $V_2\left(x\right)=0$. It should be stressed that the optimization result is not unique for the chosen boundaries. The successive runs of the optimization algorithms lead to a whole family of the potentials with very close ground state energies. This agrees with the conclusions of \cite{Rosner1978}. It is seen that there is even no qualitative agreement between the potential shown in Figure 2 (b) and the soft-core Coulomb potential (\ref{potential}) that is to be reconstructed.

\begin{figure}[h]
\centering
\includegraphics[width=.95\textwidth]{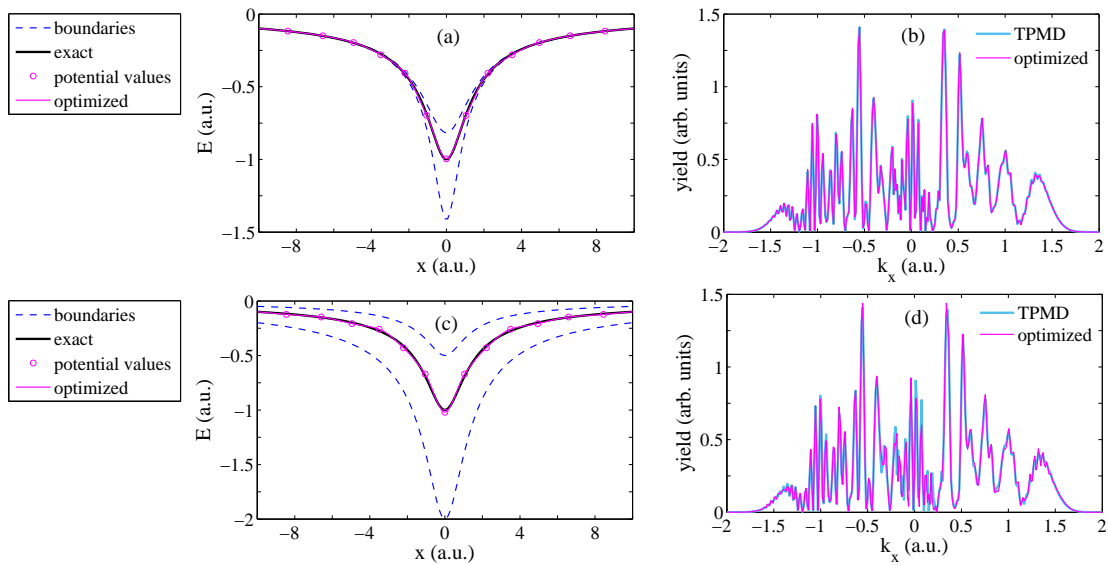} 
\caption{(a),(c) The values of the SAE potential (magenta circles) reconstructed by optimization of the electron momentum distribution, boundaries for the potential values (dashed blue curves), the potential obtained by spline interpolation using the reconstructed values (thin magenta curve), and the soft-core Coulomb potential (thick black curve). (b),(d) The PMD calculated using the optimized potential (thin magenta curve) and the TPMD (thick light-blue curve). Panels (a) and (b) show the optimization results for the case where the allowed potential values are bounded by the potentials $V_{1,2}\left(x\right)$ [equation (\ref{sandwich})] with $Z_{1,2}=1.0$ and $a_1$ and $a_2$ equal to $0.4$ and $1.5$, respectively. Panels (c) and (d) correspond to the boundaries $V_{1,2}\left(x\right)$ for the potential values given by (\ref{sandwich}) with $a_{1,2}=1.0$ and $Z_1$ and $Z_2$ equal to $0.5$ and $2.0$, respectively. The potential under reconstruction is determined by its values on the grid (\ref{nonu}) with $20$ points. The laser parameters are the same as in Figs.~\ref{fig1} and \ref{fig2}.}
\label{fig3}
\end{figure}

We now turn to the retrieval of the potential from the electron momentum distributions. At first, we use the same boundaries for the unknown potential values as in the example shown in Figure~\ref{fig2}~(a). We minimize the difference as defined by the measure (\ref{metr2}) between the PMD calculated for a potential defined by its values in the grid points and the PMD obtained for the potential (\ref{potential}) with the same laser pulse. The minimum value of the metric $\rho_2\left[{\rm PMD},{\rm TPMD}\right]$ obtained in optimization is $0.019$. The potential retrieved in the optimization procedure is shown in Figure~\ref{fig3}~(a). It is seen that the obtained potential almost coincides with the soft-core Coulomb potential (\ref{potential}) we wanted to reconstruct. The same is true for the distributions calculated for the retrieved potential and the potential (\ref{potential}), see Fig.~\ref{fig3}~(b).

The question arises how sensitive is the optimization result to a change in the boundaries $V_1\left(x\right)$ and $V_2\left(x\right)$. To answer this question, we perform another optimization with the broader boundaries for the allowed potential values. Specifically, we sandwich the potential that is to be retrieved by $V_1\left(x\right)$ and $V_2\left(x\right)$ with $a_{1,2}=1.0$ and $Z_1$ and $Z_2$ equal to $2.0$ and $0.5$, respectively. The optimization result for these broader boundaries quantitatively agrees with the potential (\ref{potential}), see Figure~\ref{fig3}~(c). The same is also true for the electron momentum distributions, see Figure~\ref{fig3}~(d). The minimum value of the metric (\ref{metr2}) obtained in optimization with these broader boundaries is $0.03$, which is only slightly higher than in the previous case. 

\begin{figure}[h]
\centering
\includegraphics[width=.95\textwidth]{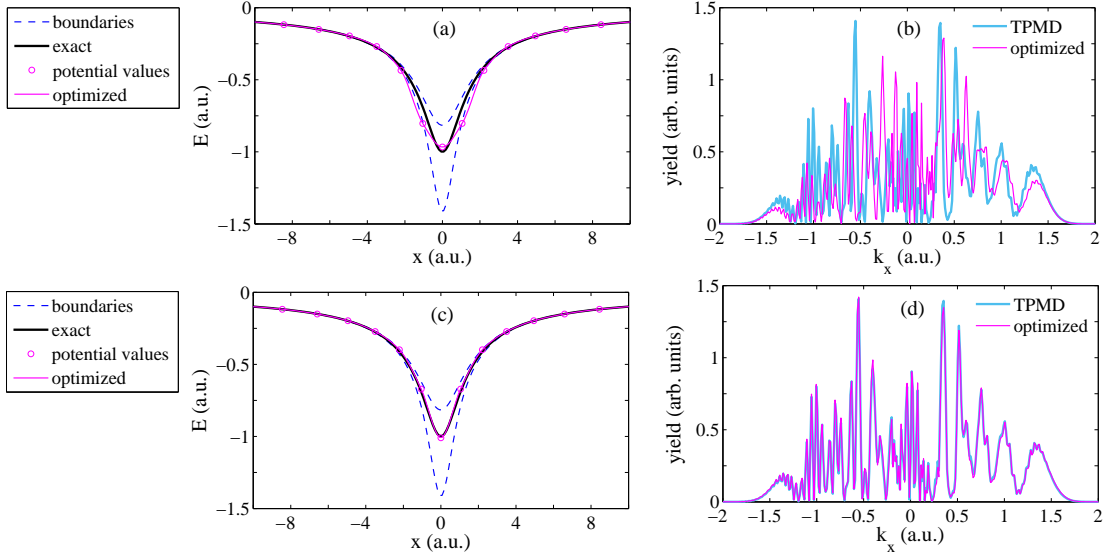} 
\caption{(a),(c) The values of the SAE potential on the non-uniform grid (\ref{nonu}) with $20$ points (magenta circles) reconstructed by optimization of the PMD, boundaries for the potential values (dashed blue curves), the potential obtained by spline interpolation based on the reconstructed values (thin magenta curve), and the soft-core Coulomb potential (thick black curve). (b),(d) The PMD calculated using the optimized potential (thin magenta curve) and the TPMD (thick light-blue curve). Panels (a) and (b) correspond to the case where the TPMD is calculated for the intensity of $2.0\times10^{14}$~W/cm$^2$ and the optimization of the momentum distributions is performed at the intensity of $1.0\times10^{14}$ W/cm$^2$. Panels (c) and (d) show the results obtained treating the field strength $F_0$ as an additional parameter that is to be optimized. The boundaries for the optimized potential values are given by equations (\ref{sandwich}) with $Z_{1,2}=1.0$ and $a_1$ and $a_2$ equal to $0.4$ and $1.5$, respectively. The parameters are the same as in Figs.~\ref{fig1}-\ref{fig3}.}
\label{fig4}
\end{figure}

We are now able to address the more important questions to the optimization-based approach, namely, how vulnerable is it to intensity fluctuations that are inevitable in an experiment? Can the actual laser intensity be restored by applying the optimization technique? In order to answer the first question, we try to retrieve the potential at the intensity of $1.0\times10^{14}$ W/cm$^2$ using the TPMD calculated for the higher intensity of $2.0\times10^{14}$ W/cm$^2$. The potential obtained in such an optimization is shown in Figure~\ref{fig4}~(a). It is seen that this potential substantially differs from the potential (\ref{potential}) we expected to reconstruct. Indeed, the minimum obtained value of the metric (\ref{metr2}) is $0.61$, and the PMD calculated for the retrieved potential does not agree with the target one, see Figure~\ref{fig4}~(b). 
To generalize our approach to the case where the TPMD is obtained at different laser intensity, we add the field strength to the parameter set that is to be optimized. This allows us to reconstruct the actual value of the laser intensity, at which the TPMD that we want to reproduce in optimization is obtained. In this case, the surrogate optimization turns out to be slightly more efficient than the particle swarm method. The results of this modified approach are shown in Figures~\ref{fig4}~(c) and (d). It is seen from Figure~\ref{fig4}~(c) that the retrieved potential is in a quantitative agreement with the soft-core Coulomb potential (\ref{potential}) used to calculate the TPMD. The same is true for the resulting electron momentum distribution, see Figure~\ref{fig4}~(d). The difference between the momentum distributions $\rho_2\left[{\rm PMD},{\rm TPMD}\right]$ comes to only $0.032$. The optimized value of the field strength is $0.0763$~a.u., whereas the exact value of $F_0$ equals to $0.0755$~a.u. Although these results are encouraging, it is clear that further studies are needed to completely tackle the question regarding the intensity fluctuations. To mimic a real experimental situation, it is necessary to average the PMD over the intensity distribution within the focal volume at every iteration of the optimization process. We leave this modification of the proposed method for further studies implying the generalization to the 3D case.

\begin{figure}[h]
\centering
\includegraphics[width=.95\textwidth]{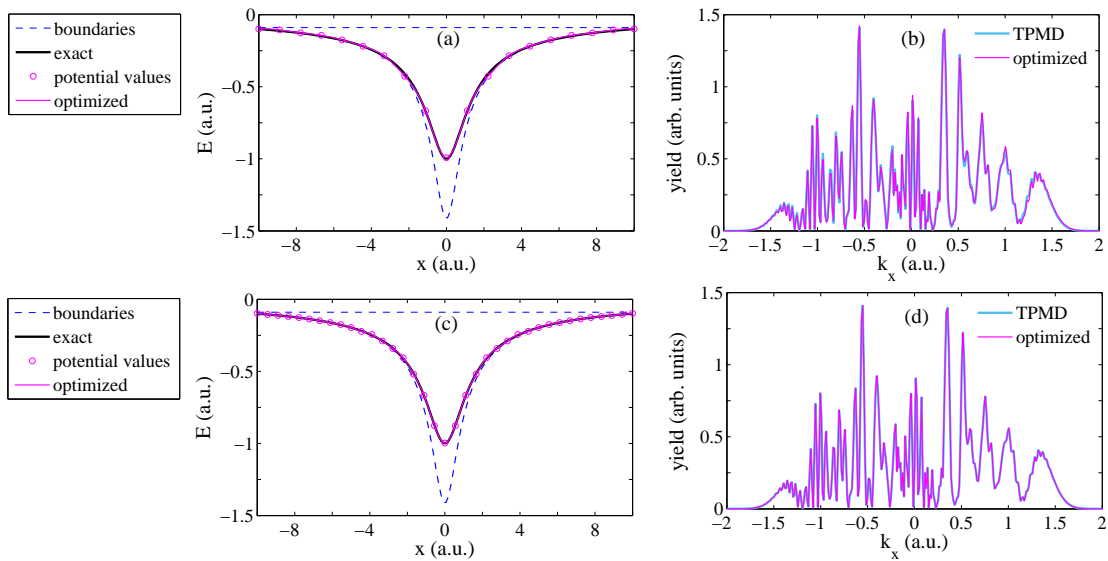} 
\caption{Panels (a) and (b) show the same as Figures~\ref{fig3}~(a) and (b) for the optimization parameters varying in wider ranges and the potential determined by the values on the uniform grid consisting of $10$ points between $0$ and $10.0$ a.u. Panels (c) and (d) display the optimization results for the same grid consisting of $20$ points. The optimization parameters are bounded by the potential $V_1\left(x\right)$ calculated from equation (\ref{sandwich}) with $Z_1=1.0$ and $a_1=0.4$ and $V_{2}\left(x\right)=-0.09$~a.u. The parameters are the same as in Figures~\ref{fig1}-\ref{fig3}.}
\label{fig5}
\end{figure}

The results shown in Figures \ref{fig2}-\ref{fig4} were obtained for the non-uniform grid (\ref{nonu}) in the range $\left[-x_{max}, x_{max}\right]$. However, the typical molecular potential we intend to reconstruct has a long-range Coulomb asymptotic at large distance: $V\left(r\right)\rightarrow 1/r$ at $r\rightarrow \infty$. Therefore, there is no need to find the values of the potential at large $x$, and an optimization of the potential values at $x\rightarrow \infty$ leads to a waste of computational resources. At the same time, it is highly desirable to have a denser grid for small $x$, where the potential can vary significantly. To address both these issues, from this point on we use a smaller computational box $\left[-x_{C}, x_{C}\right]$ setting $V\left(x\right)=1/\left|x\right|$ for $\left|x\right|>x_{C}$. Typically, we chose $x_{C}=10.0$~a.u. and use uniform grid within the range $\left[-x_{C}, x_{C}\right]$. The optimization results are shown in Figures~\ref{fig5}~(a) and (b). It is seen from Figure~\ref{fig5}~(a) that retrieved potential agrees well with the soft-core Coulomb potential (\ref{potential}). The optimization results in a measure (\ref{metr2}) equal to $0.04$ what corresponds to a good agreement between the obtained and target electron momentum distributions, see Figure~\ref{fig5}~(b). A different optimization method was used here. At first, we performed an optimization of the bound state and first excited state energies. As mentioned above, such optimization results in a family of different potentials, unless the boundaries for the potential values are close to each other. Some of these potentials resemble the desired potential (\ref{potential}), whereas the others are very different from it and do not match the Coulomb asymptote at $x=\pm x_{C}$. Then we use all the obtained potentials as an initial approximation for the second optimization procedure that minimize the difference between the corresponding PMD and the target one. For the second optimization we use the pattern search algorithm (Hook-Jeeves method \cite{HookeJeeves1961}), see, e.g., \cite{Kolda2003, Audet2003} for reviews. This combined two-step approach to optimization of the PMD turns out to be computationally more efficient than particle swarm optimization we used before and many other gradient-free optimization methods, including simulated annealing, genetic algorithms, and surrogate optimization, if these methods start from a random initial approximation.  

The optimization results presented in Figures~\ref{fig5}~(a) and (b) were obtained for $20$ grid points in the range $\left[-x_{C}, x_{C}\right]$ (i.e, for $10$ grid points for $x\in\left[0, x_{C}\right]$). This corresponds to the grid spacing $dx=1.0$~a.u. Suppose that we need better resolution along the $x$ axis, e.g., $dx=0.5$~a.u., what corresponds to $20$ grid points between $0$ and $x_{C}$. The most efficient way to perform the optimization of the PMD on a denser grid is to use the results obtained for a sparser grid as an initial approximation. In doing so the potential values in the points of a denser grid can be restored by interpolation. The results of the application of this approach are shown in Figures~\ref{fig5}~(c) and (d). Note that for $dx=0.5$~a.u. we achieve a perfect agreement between the retrieved potential and the desired soft-core Coulomb potential (\ref{potential}). The corresponding PMD's are also almost indistinguishable from each other, see Figure~\ref{fig5}~(d). The optimization algorithm terminates when the distance $\rho_2\left[{\rm PMD},{\rm TPMD}\right]$ reduces to $0.0042$.

\subsection{Reconstruction of parametrized potential}

Up to this point we have not used any parametrization to represent the unknown SAE potential. It is clear, however, that this parametrization-free approach being extended to the two-dimensional and especially the 3D case will become a very difficult computational task. Indeed, such an extension will lead to a necessity of optimization of a function that depends on hundred of variables. We note that this task is feasible with modern computational facilities and optimization algorithms. Such problems arise, for example, in research on magnetically confined plasmas for fusion energy, see, e.g. \cite{Spong2001, Drevlak2019}. It is nevertheless of interest to test the optimization-based algorithm in the case where the unknown potential is in some way parametrized. It is natural to express the SAE potential as
\begin{equation}
V\left(x\right)=V_{0}\left(x\right)\left[1+V_{1p}\left(x\right)\right],
\label{repr}
\end{equation}
where $V_{0}\left(x\right)$ is a known potential with correct asymptotic behavior and the potential $V_{1}\left(x\right)$ is to be parametrized and determined. Here we choose $V_{0}\left(x\right)=Z_{0}/\left(x^4+a_0\right)^{\frac{1}{4}}$ with $Z_{0}=1.0$ and $a_{0}=1000.0$ what corresponds to a very shallow and wide potential well, which is substantially different from the potential (\ref{potential}) that should be restored. The question thus arises how to parametrize the potential $V_{1}\left(x\right)$ in the best possible way.

Since we know the potential $V\left(x\right)$ we want to reconstruct, this question is easy to answer. Indeed, by trying different options and applying standard curve fitting routines \cite{Matlab}, we find that the rational interpolation can be used to approximate the function $V_{1p}\left(x\right)=V\left(x\right)/V_{0}\left(x\right)-1$ with a good accuracy. Therefore, the potential $V_{1p}\left(x\right)$ can be represented as a quotient of two polynomials $P_m\left(x\right)$ and $Q_{n}\left(x\right)$:
\begin{equation}
V_{1p}\left(x\right)=\frac{P_{m}\left(x\right)}{Q_{n}\left(x\right)}
\label{ratrep}
\end{equation}
It is clear that since $V_{1p}\left(x\right)\to0$ at $x\to0$, the inequality $m<n$ should be fulfilled. It is easy to see that the measure $\rho_2$ of the difference between the corresponding momentum distribution and the target one dramatically changes for different choices of the $m$ and $n$. If, for example, $m=0$ and $n=2$, the minimum value of the $\rho_2$ that can be achieved is equal to $0.29$. This corresponds to $V_{1p}\left(x\right)=p_{1}/\left(x^{2}+q_1\right)$ with $p_1=29.99$ and $q_1=3.62$. The linear function in the nominator $\left(m=1\right)$ and the quadratic function in the denominator $\left(n=2\right)$, i.e., $V_{1p}\left(x\right)=\left(p_{1}x+p_{2}\right)/\left(x^2+q_{1}x+q_{2}\right)$, allow to reduce the measure (\ref{metr2}) to $0.047$. This value is achieved for $p_{1}=-1.15$, $p_{2}=36.90$, $q_{1}=0.92$, and $q_{2}=23.95$. The minimum value of the $\rho_2$ we have obtained using the equation (\ref{ratrep}) is equal to $0.01$. It corresponds to the constant in the numerator and the fourth-order polynomial in the denominator of the quotient (\ref{ratrep}):
\begin{equation}
V_{1p}\left(x\right)=\frac{p_{1}}{x^4+q_{1}x^3+q_{2}x^2+q_{3}x+q_{4}},
\label{four}
\end{equation}
where $p_{1}=1261.0$, $q_{1}=-9.747$, $q_{2}=66.5$, $q_{3}=10.4$, and $q_{4}=138.7$. It should be noted that not all of the coefficients in the denominator of this formula are positive. When using the optimization-based method in practice, a series of optimizations applying different ways to parametrize the unknown potential should be performed. This will allow to compare the optimization results and to choose the best way of the parametrization similarly to what we do here with the fitting of the known potential. 

\begin{figure}[h]
\centering
\includegraphics[width=.95\textwidth,trim={6cm 0mm 0mm 0mm}]{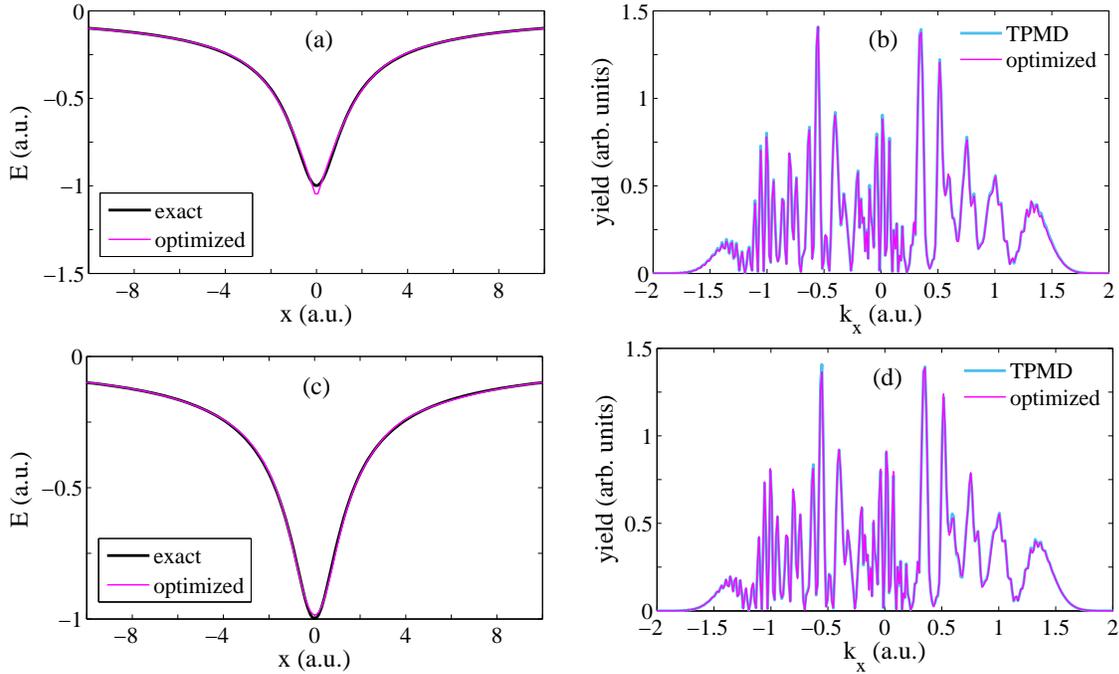} 
\caption{Optimization results for parametrized SAE potential. Panels (a) and (c) show the reconstructed potentials (thin magenta curve) and the soft-core Coulomb potential (thick black curve). Panels (b) and (d) display the comparison of the PMD's obtained from the TDSE with the retrieved (thin magenta curve) and the exact (thick light-blue curve) SAE potentials. Panels [(a), (b)] and [(c), (d)] correspond to the parametrization (\ref{four}) and (\ref{expret}), respectively. The laser parameters are the same as in Figs.~\ref{fig1}-\ref{fig3}.}
\label{fig6}
\end{figure}

We now try to recover the parameter values in the parametrization (\ref{four}) by optimizing the PMD. At first glance it would seem that optimization of function (\ref{four}) depending on only $5$ parameters is a simple task compared to the one performed in Sec.~3.1. But this is not the case, since negative values of the parameters $q_{i} \left(i=1,...,4\right)$ may lead to $1+V_{1p}\left(x\right)<0$ and, therefore, $V\left(x\right)>0$ for certain ranges of $x$. To prevent this situation, we use constrained optimization. Specifically, instead of minimizing the measure (\ref{metr2}) alone, we now look for a minimum of
\begin{equation}
\rho_2\left[{\rm PMD},{\rm TPMD}\right]+w \cdot V_{m}^{2}\left[1+{\rm sgn}\left(V_m\right)\right],
\label{constrained}
\end{equation}
where $V_m$ is the maximum value of $V\left(x\right)$ in the interval $\left[0, x_0\right]$, and $w$ is some weight factor. Typically, we use $w$ in the range between $5.0$ and $20.0$. We allow for the following ranges of the optimization parameters: $p_1\in\left[-2000.0, 2000.0\right]$ and $q_{i}\in\left[-100.0, 100.0\right] \left(i=1,...,4\right)$. To speed up the simulations, we again perform the optimization in two steps, i.e., we use the results obtained in optimization of the ground state, first, and second excited states energies as initial approximations for the optimization of the PMD. The retrieved potential and the corresponding electron momentum distribution are shown in Figures~\ref{fig6}~(a) and (b), respectively. The following parameter values were obtained: $p_{1}$=403.44, $q_{1}=3.08$, $q_{2}=17.03$, $q_{3}=35.57$, and $q_{4}=80.45$. The difference $\rho_2$ between the resulting PMD and the TPMD is $0.0164$, what is higher than the one corresponding to the parameters obtained by approximation of the known dunction $V_{1p}\left(x\right)$. Therefore, some local minimum, albeit quite close to the desired global one, is found in optimization. It is seen that the retrieved potential agrees well with the exact result, see Figure~(9)~(a). The same is also true for the electron momentum distributions [see Figure \ref{fig6}~(b)].
As a next step in testing the method, we represent our potential as a sum of a few Gaussian functions:
\begin{equation}
V_{1p}\left(x\right)=\sum_{k=1}^{k_{max}}a_{k}\exp\left[-\frac{\left(x-b_{k}\right)^2}{c_{k}^{2}}\right]
\label{expret}
\end{equation}
The representation (\ref{expret}) is obviously more flexible, i.e., it allows for finer variations of the potential function, as compared to the one applying rational function. Here we choose $k_{max}=5$ what corresponds to $15$ parameters to be optimized. It should be stressed that in contrast to all other examples shown in this paper, here we do not assume any symmetry of the potential $V\left(x\right)$. Indeed, the allowed values of the optimization parameters are: $a_{k}\in\left[-4.0, 4.0\right]$, $b_{k}\in\left[-20.0, 20.0\right]$, and $c_{k}\in\left[0, 100.0\right]$ $\left(k=1,...,5\right)$. As an initial approximation for the optimization of the PMD's we use the potential shown in Figure~8~(a), i.e., the one obtained with parametrization (\ref{ratrep}). The optimization results are presented in Figures \ref{fig6}~(c) and (d). As expected, a better agreement between the retrieved potential and the exact one can be achieved with parametrization (\ref{expret}). However, the measure (\ref{metr2}) remains practically unchanged: $\rho_2\left[{\rm PMD},{\rm TPMD}\right]=0.0159$. 

The examples above demonstrate that in the 1D case the parametrization of the potential does not offer any decisive advantages compared to the direct representation on a grid. Nevertheless, it is shown that the optimization-based method also works in the case, where the potential is determined by a number of parameters. This is essential in view of the possible extension of the approach onto the 3D case, where parametrization of the unknown potential is expected to become particularly important.

\section{Conclusions and outlook}

In conclusion, we have developed a method capable to retrieve the SAE potential in an atom or molecule from a given momentum distribution of photoelectrons ionized by a strong laser pulse. In this method the potential is found by minimization of the difference between the given momentum distribution and the one obtained from the solution of the TDSE with the SAE potential that varies in the optimization process. The unknown potential is either represented by a set of its values in points of a spatial grid, or by a set of parameters. We have shown that the optimization can be performed using a number of different derivative-free techniques, including particle swarm method, surrogate optimization, and pattern search. It is found that the most efficient approach is based on the use of potentials obtained in optimization of a few bound-state energies as initial approximations for the optimization of the PMD.

We have tested our method by reconstructing of the soft-core Coulomb potential from the corresponding PMD generated in ionization of a 1D atom by a strong few-cycle laser field. It is shown that the retrieved SAE potential is in a quantitative agreement with the potential we aimed to reconstruct. This is true for both ways used to represent the potential under reconstruction. In the case where the potential is represented by its values on a grid the spatial resolution can be effectively improved by using the optimization results on a sparse grid as an initial approximation for optimization on the dense grid. This allows to avoid severe computational costs when optimizing a function depending on a few dozens of variables.

It is clear that the measured electron momentum distributions are affected by focal averaging. We have shown that the actual laser intensity can be restored together with the SAE potential in the optimization approach. Nevertheless, further work is needed to fully explore the question how sensitive is the proposed method to the focal averaging. Furthermore, extension of the method to the real 3D case require a reliable measure used to compare different momentum distributions. To this end, the tools of image analysis and pattern recognition can be applied. It remains to be studied which of these tools are the most appropriate for the problem at hand. Finally, we have restricted ourselves by the optimization of only one part of the PMD created by the direct electrons. However, our preliminary results show that the momentum distribution of the rescattered photoelectrons are more sensitive optimization target that can be used for the retrieval of the potential. It is thus of interest to develop a method that optimizes the distributions of the rescattered electrons. Therefore, future work is needed to address the issues listed here. Developments in these directions have already begun. We believe that the advent of the method for retrieval of the SAE potential from the electron momentum distribution will be an important step forward in the studies of strong-field ionization.  

\ack 
We are grateful to Professor~Manfred~Lein, Florian~Oppermann, Simon~Brennecke, and Shengjun Yue for valuable discussions and continued interest to this work. This work was supported by the Deutsche Forschungsgemeinschaft (Grant No. SH 1145/1-2).


\section*{References}
\begin{thebibliography}{99}
\bibitem{Marangos2004} Marangos J P 2004 Molecules in a strong laser field ({\it Atoms and Plasmas in Super-Intense Laser Fields, SIF Conference Proceedings}) ed D~Batani, C~J~Joachain, and S~Martellucci \textbf{88} 213 (Societ{\`a} Italiana di Fisica, Bologna)

\bibitem{Bandrauk2003} Bandrauk A D and Kono H 2003 Molecules in intense laser fields: nonlinear multiphoton spectroscopy and near-femtosecond to sub-femtosecond (attosecond) dynamics ({\it Multi-Photon Processes and Spectroscopy}) ed S~H~Lin, A~A~Villaeys, and Y Fujimura \textbf{15} 147 (World Scientific, Singapore)

\bibitem{BeckerRev2002} Becker~W, Grasbon~F, Kopold~R, Milo\v{s}evi\'c~D~B, Paulus~G~G, and Walther~H 2002 Above-threshold ionization: From classical features to quantum effects Adv. At. Mol. Opt. Phys. \textbf{48} 35

\bibitem{MilosevicRev2003} Milo\v{s}evi\'c~D~B and Ehlotzky~F 2003 Scattering and reaction processes in powerful laser fields Adv. At. Mol. Opt. Phys. \textbf{49} 373    

\bibitem{FaisalRev2005} Becker A and Faisal~F~H~M 2005 Intense field many-body S-matrix theory, J. Phys. B: At. Mol. Opt. Phys. \textbf{38} R1 

\bibitem{FariaRev2011} Faria~C and Liu~X 2011 Electron-electron correlation in strong laser fields J. Mod. Opt. \textbf{58} 1076

\bibitem{Muller1999} Muller~H~G 1999 An efficient propagation scheme for the time-dependent Schr\"dinger equation in the velocity gauge Las. Phys. \textbf{9} 138

\bibitem{Bauer2006} Bauer~D and Koval~P 2006 Qprop: A Schr\"{o}dinger-solver for intense laser–atom interaction Comput. Phys. Comm. \textbf{174} 396 

\bibitem{Madsen2007} Madsen~L~B, Nikolopoulos~L~A~A, Kjeldsen~T~K, and Fern\'andez~J 2007 Extracting continuum information from $\Psi$(t) in time-dependent wave-packet calculations Phys. Rev. A \textbf{76} 063407 

\bibitem{Tong2017} Tong~X~M 2017 A three-dimensional time-dependent Schr\"{o}dinger equation solver: An application to hydrogen atoms in an elliptical laser field J. Phys. B. \textbf{50} 144004

\bibitem{Keldysh1964} Keldysh~L~V 1964 Ionization in the field of a strong electromagnetic wave Sov. Phys. JETP \textbf{20} 1307 

\bibitem{Faisal1973} Faisal~F~H~M 1973 Multiple absorption of laser photons by atoms, J. Phys. B.: At. Mol. Opt. Phys. \textbf{6} L89

\bibitem{Reiss1980} Reiss~H~R 1980 Effect of an intense electromagnetic field on a weakly bounded system Phys. Rev. A \textbf{22} 1786

\bibitem{Linden1988} van Linden van~den~Heuvell~H~B and Muller~H~G 1988 in \textit{Multiphoton processes, Studies in Modern Optics}, ed S~J~Smith and P~L~Knight (Cambrige University, Cambrige)

\bibitem{Gallagher1988} Gallagher~T~F 1988 Above-Threshold ionization in low-frequency limit Phys. Rev. Lett. \textbf{61} 2304

\bibitem{Corkum1989} Corkum~P~B, Burnett~N~H, and Brunel~F 1989 Above-threshold ionization in the long-wavelength limit Phys. Rev. Lett. \textbf{62} 1259 

\bibitem{Kulander_Schafer1993} Kulander~K~C, Schafer~K~J, and Krause~J~L 1993 in \textit{Super-Intense Laser-Atom Physics} ed B~Pireaux, A~L'Hullier and K~Rzazewski (Plenum, New York)

\bibitem{Corkum1993} Corkum~P~B 1993 Plasma perspective on strong-field multiphoton ionization Phys. Rev. Lett. \textbf{71} 1994

\bibitem{Palacios2006} Palacios~A, Bachau~H and Mart\'{\i}n~F 2006 Enhancement and Control of ${\mathrm{H}}_{2}$ Dissociative Ionization by Femtosecond VUV Laser Pulses Phys. Rev. Lett. \textbf{96} 143001 

\bibitem{Saenz2014} F\"orster~J, Vanne~Y~V, and Saenz~A 2014 Ionization behavior of molecular hydrogen in intense laser fields: Influence of molecular vibration and alignment Phys. Rev. A 90 053424

\bibitem{Kulander1988} Kulander~K~C 1988 Time-dependent theory of multiphoton ionization of xenon Phys. Rev. A \textbf{38} 778  

\bibitem{Kulander1991} Kulander~K~C, Schafer~K~J, and Krause~J~L 1991 Single-active electron calculation of multiphoton process in krypton, Int. J. Quantum Chem. \textbf{25} 415 

\bibitem{Gamess} Gordon~M~S, Jensen~J~H, Koseki~S, Matsunaga~N, Nguyen~K~A, Su~S, Windus~T~L, Dupuis~M, and Montgomery~J~A 1993 J. Comput. Chem. General atomic and molecular electronic structure system \textbf{14} 1347

\bibitem{Gaussian} Frisch~M~J et al. 2016 \textit{GAUSSIAN 16 Revision C.01} (Gaussian, Inc. Wallingford, CT) 

\bibitem{KohnSham1965} Kohn~W and Sham~L~J 1965 Self-Consistent Equations Including Exchange and Correlation Effects Phys. Rev. \textbf{140} A1133

\bibitem{vanLeeuwen1994} van Leeuwen~R and Baerends~E~J 1994 Exchange-correlation potential with correct asymptotic behavior Phys. Rev A \textbf{49} 2421

\bibitem{Madsen2010} Abu-samha~M and Madsen~L~B 2010 Single-active-electron potentials for molecules in intense fields Phys. Rev A \textbf{81} 033416

\bibitem{Lin2010} Zhao~S~F, Jin~C, Le~A-T, Jiang~T~F, and Lin~C~D 2010 Determination of structure parameters in strong-field tunneling ionization theory of molecules Phys. Rev A \textbf{81} 033423. 

\bibitem{Miller1974} Miller~K~J and Green A~S~E 1974 Energy levels and potential energy curves for H$_2$, N$_2$, and O$_2$ with an independent particle model J. Chem. Phys \textbf{60} 2617 

\bibitem{Kastner2010} K\"astner A, Grossmann~F, and Schmidt~R 2010 Reliability of soft-core approximations in theoretical studies of molecules in intense laser fields Phys. Rev. A \textbf{81} 023414

\bibitem{Atabek2012} Peters~M, Nguyen-Dang~T~T, Charron~E, Keller~A, and Atabek~O 2012 Laser-induced electron diffraction: A tool for molecular orbital imaging Phys. Rev. A \textbf{85}, 053417

\bibitem{Ding2016} Lv~H, Zuo~W, Zhao~L, Xu~H, Jin~M, and Ding~D 2016 Comparative study on atomic and molecular Rydberg-state excitation in strong infrared laser fields Phys. Rev. A \textbf{93} 033415

\bibitem{WerschnikGross2007} Werschnik~J and Gross~E~K~U 2007 Quantum optimal control theory J. Phys. B: At. Mol. Opt. Phys. \textbf{40} R175

\bibitem{Rabitz2010} Brif~C, Chakrabarti~R, and Rabitz~H 2010 Control of quantum phenomena: past, present and future New J. Phys. \textbf{12} 075008  

\bibitem{Castro2009} Castro~A, R\"as\"anen~E, Rubio~A, and Gross~E~K~U 2009 Femtosecond laser pulse shaping for enhanced ionization Europhys. Lett. \textbf{87} 53001  

\bibitem{Solanpaa2014} Solanp\"a\"a~J, Budagosky~J~A, Shvetsov-Shilovski~N~I, Castro~A, Rubio~A, and R\"as\"anen~E 2014 Optimal control of high-harmonic generation by intense few-cycle pulses Phys. Rev. A \textbf{90} 053402

\bibitem{Shvetsov2015} Shvetsov-Shilovski~N~I, Madsen~L~B, and R\"as\"anen~E 2015 Suppression of strong-field ionization by optimal pulse shaping: Application to hydrogen and the hydorgen molecular ion Phys. Rev. A \textbf{91} 023425

\bibitem{Koch2016} Goetz~R~E, Karamatskou~A, Santra~R and Koch~C~P 2016 Quantum optimal control of photoelectron spectra and angular distributions Phys. Rev A \textbf{93} 013413

\bibitem{Wang2004} Wang~Z, Bovik~A~C, Sheikh~H~R, and Simoncelli~E~P 2003 Image quality assessment: from error visibility to structure similarity IEEE Trans. Image Process \textbf{13} 600

\bibitem{Buldas2013} Buldas~A, Kronmaa~A, and Laanoja~R 2013 {\it Keyless Signature's Infrastructure: How to Build Global Distributed Hash-Trees} (Berlin, Heidelberg: Springer)

\bibitem{KolmogorovFomin} Kolmogorov A N, Fomin S V, and Silverman R A 1975 {\it Introductory real analysis} (New York: Dover Publications)

\bibitem{Anderson1999} Anderson E et al 1999 {\it LAPACK User's Guide}, 3rd ed. (Philadelphia: Society for Industrial and Applied Mathematics)

\bibitem{Kjeldsen2007} Kjeldsen T K 2007 {\it Wave packet dynamics studied by ab initio methods: Application to strong-field ionization of atoms and molecules} Ph.D. thesis (\AA rhus: University of \AA rhus, Denmark) 

\bibitem{BauerBook} Bauer D (ed) 2017 {\it Computational Strong-Field Quantum Dynamics} (Berlin/Boston: Walter de Gruyter GmbH)

\bibitem{Feit1982} Feit M D, Fleck J A, and Steiger A 1982 Solution of the Schr\"{o}dinger equation by a spectral method J. Comput. Phys. \textbf{47} 412

\bibitem{Tong2006} Tong X M, Hino K, and Toshima N 2006 Phase-dependent atomic ionization in few-cycle intense laser pulse Phys. Rev. A \textbf{74} 031405(R)

\bibitem{Rios2013} Rios~L~M and Sahinidis~N~V 2013 Derivative-free optimization: a review of algorithms and comparison of software implementations J. Global Optim. \textbf{56} 1247 

\bibitem{Wild2019} Larson~J, Menickelly~M, and Wild~S~M 2019 Derivative-free optimization methods Acta Numer. \textbf{28} 287

\bibitem{Kennedy1995} Kennedy~J and Eberhart~R 1995 Particle Swarm Optimization {\it Proc. of IEEE Int. Conf. on Neural Networks} vol~4 p~1942 

\bibitem{Shi1998} Shi~Y and Eberhart~R~C 1998 A modified particle swarm optimizer {\it Proc. of IEEE Int. Conf. on Evolutionary Computation} p~69 
 
\bibitem{Gutmann2001} Gutmann~H-M 2001 A radial basis function method for global optimization. J. Glob. Optim \textbf{19} 201

\bibitem{HookeJeeves1961}  Hooke R and Jeeves T A 1961 "Direct search" solution of numerical and statistical problems J. ACM. \textbf{8}(\textbf{2}) 212

\bibitem{Kolda2003} Kolda~T~G, Lewis~R~M, and Torczon~V 2003 V Optimization by direct search: new perspectives on some classical and modern methods SIAM Rev. \textbf{45} 385

\bibitem{Audet2003} Audet~C and Dennis~J~E Jr 2003 Analysis of Generalized Pattern Searches SIAM J. Optimiz. \textbf{13} 889

\bibitem{Matlab} MATLAB 2020 version 9.9.0 (R2020b) Natick, Massachusetts: The MathWorks Inc

\bibitem{Eberly1988} Javanainen~J, Eberly~J~H, and Su~Q 1988 Numerical simulations of multiphoton ionization and above-threshold electron spectra Phys. Rev. A \textbf{38} 3430 

\bibitem{Press1992} Press~W~H, Teukolsky~S~A, Vetterling~W~T and Flannery~B~P 1992 {\it Numerical Recipes in Fortran 77: The Art of Scientific Computing} 2nd ed (Cambridge: Cambridge University Press) 

\bibitem{Wang1994} Wang~J, Chu~S~I, and Laughlin~S 1994 Multiphoton detachment of ${\mathrm{H}}^{\mathrm{\ensuremath{-}}}$. II. Intensity-dependent photodetachment rates and threshold behavior---complex-scaling generalized pseudospectral method Phys. Rev. A \textbf{50} 3208

\bibitem{Yao1993} Yao~G and Chu~S~I 1993 Generalized pseudospectral methods with mappings for bound and resonance state problems Chem. Phys. Lett. \textbf{204} 381

\bibitem{Chu1997} Tong~X-M and Chu~S~I 1997 Theoretical study of multiple high-order harmonic generation by intense ultrashort laser fields: A new generalized pseudospectral time-dependent method Chem. Phys. \textbf{217} 119. 

\bibitem{Rosner1978} Thacker~H~B, Quigg~C, and Rosner~J~L 1978 Inverse scattering problem for quarkonium systems. II. Applications to $\ensuremath{\psi}$ and $\ensuremath{\Upsilon}$ families Phys. Rev D \textbf{18} 287

\bibitem{Spong2001} Spong~D~A, Hirshman~S~P, Berry~L~A, Lyon~J~F, Fowler~R~H, Strickler~D~J, Cole~M~J, Nelson~B~N, and Williamson~D~E (2001) Physics issues of compact drift optimized stellarators Nucl. Fusion \textbf{41} 711

\bibitem{Drevlak2019} Drevlak~M, Beidler~C~D, Geiger~J, Helander~P, and Turkin~Y 2019 Optimization of stellarator equilibria with ROSE Nuclear Fusion \textbf{59} 016010 

\end{thebibliography}
\end{document}